\documentclass[showpacs,nofootinbib]{revtex4}
\usepackage{eurosym}
\usepackage{amsfonts}
\usepackage{amsmath}
\usepackage{graphicx}
\usepackage{amssymb}

\input{tcilatex}
\begin{document}

\title{An Investigation of Equivalence between the Bulk-based and the
Brane-based Approaches for Anisotropic Models}
\author{G\"{u}l\c{c}in Uluyaz\i }
\affiliation{Physics Department, \.{I}stanbul University, Vezneciler, Turkey.}

\begin{abstract}
We investigate the relation between the brane-based and the bulk-based
approaches for anisotropic case in brane-world models. In the brane-based
approach, the brane is chosen to be fixed on a coordinate system, whereas in
the bulk-based approach it is no longer static as it moves along the extra
dimension. It was shown that these two approaches are basicly equal for
specific models \cite{Ref1}, \cite{Ref2}. In this paper, it is aimed to get
general formalism of the equivalence obtained in Mukohyama et al. \cite{Ref1}%
. We found that calculations driven by a general anisotropic bulk-based
metric yield a brane-based metric in Gaussian Normal Coordinates by
conserving spatial anisotropy. We also derive solutions for an anisotropic
bulk-based model and apply to get corresponding brane-based metric of the
model.
\end{abstract}

\date{17.12.2011}
\maketitle

\section{Introduction}

Although extra dimensions have been widely involved in the String Theory for
several decades, their introduction into cosmology has been recently offered
to give provide a new perception of the Universe and its evolution. It was
proposed \cite{Ref3} that our Universe may be a 3- dimensional surface
(called domain wall or brane) embedded in a higher dimensional space (called
bulk). Following the novel approach to hierarchy problem proposed by
Arkani-Hamed, Dimopoulos and Dvali \cite{Ref4}-\cite{Ref6}, Rundall and
Sundrum \cite{Ref7},\cite{Ref8} suggested that the brane representing our
universe lies in a 5-dimensional anti-de-Sitter (AdS) bulk, which is
strongly curved in order to ensure that gravitation has effectively 4-
dimensional character on the brane, even if the extra dimension is finite.

The key feature of the Randall Sundrum (RS) models is that the induced
metric on the brane is flat Minkowski spacetime. The brane spacetime is
generally static owing to the effect of brane energy and the tension is
neutralized in the presence of bulk cosmological constant. However our
universe is not Minskowskian and has an expansion. To allow time dependent
expansion along spatial directions in a brane-world model, some authors \cite%
{Ref9} - \cite{Ref20} consider the "brane-based approach" in which the brane
is fixed at a point along the extra dimension. Although the cosmology of the
brane is more apparent in this approach, bulk spacetime structure remains
less transparent.

The brane based approach is usually handled in Gaussian Normal Coordinates,
in which the metric component of extra dimension is normal to the brane.
Writing 5-dimensional Einstein Field Equations 
\begin{equation}
R_{AB}^{5}-\frac{1}{2}g_{AB}^{5}=-\Lambda _{5}g_{AB}^{5}+\kappa
_{5}^{2}T_{AB}^{5}
\end{equation}%
together with Gauss-Codacci equations

\begin{equation}
R_{ABCD}^{4}=h_{A}^{E}h_{B}^{F}h_{H}^{G}h_{D}^{E}R_{EFGH}^{5}+K_{AC}K_{BD}-K_{AD}K_{BC}
\end{equation}%
\begin{equation}
\nabla _{B}^{4}K_{A}^{B}-\nabla _{A}^{4}K=h_{A}^{B}R_{BC}^{5}n^{C}
\end{equation}%
and Israel Junction Conditions, 
\begin{equation}
\left[ h_{AB}\right] =0
\end{equation}%
\begin{equation}
\left[ K_{AB}\right] =\kappa _{5}^{2}\left( S_{AB}-\frac{1}{3}h_{AB}S\right)
\end{equation}%
cosmological properties of brane can be found. Here ${T_{AB}^{5}}$
represents any 5-dimensional energy-momentum of the gravitational sector and
it provides a conservation equation $\nabla _{A}T^{AB}=0$. ${S_{AB}}$ and ${%
K_{AB}}$ represent energy momentum tensor of brane and extrinsic curvature
tensor respectively, so that ${S}$ and ${K}$ are their traces. ${n^{A}}$
being the unit vector normal to the brane, $h_{AB}=g_{AB}-n_{A}n_{B}$ is the
induced metric on the brane.

On the other hand, using induced field equations is a more elegant way for
the derivation of the brane world cosmological equations.
Shiromizu-Maeda-Sasaki \cite{Ref21} obtained 4-dimensional induced field
equations by projecting 5-dimensional equations onto the brane. 
\begin{equation}
G_{AB}^{4}=-\Lambda _{4}h_{AB}+8\pi G_{5}\tau _{AB}+\kappa _{5}^{4}\pi
_{AB}^{5}-E_{AB}^{5}
\end{equation}%
\begin{equation}
\Lambda _{4}=\frac{1}{2}\kappa _{5}^{2}[\Lambda _{5}+\frac{1}{2}\kappa
_{5}^{2}\lambda ^{2}]
\end{equation}%
\begin{equation}
G_{5}=\frac{\kappa _{5}^{4}\lambda }{48\pi }
\end{equation}%
\begin{equation}
\pi _{AB}^{5}=-\frac{1}{4}\tau _{AC}{\tau _{B}}^{C}+\frac{1}{12}\tau \tau
_{AB}+\frac{1}{8}h_{AB}\tau _{CD}\tau ^{CD}-\frac{1}{24}h_{AB}\tau ^{2}
\end{equation}%
These equations are also important to include bulk effects in the
4-dimensional spacetime.

Apart from the brane-based approach, the bulk-based approach accepts a
moving brane (or a domain wall) following some timelike trajectories in a
static bulk spacetime. It was first obtained by Ida \cite{Ref22} in which
the most general static AdS solution 
\begin{equation}
ds^2=h(r)dt^2-h^{-1}(r)dr^2-r^2\left[\frac{d\chi^2}{1-\kappa\chi^2}%
+\chi^2d\Omega_{II}^2\right]
\end{equation}
where $h(r)=\kappa-\frac{\mu}{r^2}+k^2r^2$.

The bulk-based approach is quite general and does not depend on $Z_2$
symmetry around the brane. The cosmological solution of the brane will be
described by its movement in the bulk. Some other papers using this approach
are given in \cite{Ref23}-\cite{Ref27}.

It was found by Mukohyama et. al. \cite{Ref1} that these two approaches are
equivalent. They have constructed a coordinate transformation relating Ida's
solution \cite{Ref22} to Binetruy et. al. \cite{Ref9}. In contrast with \cite%
{Ref1}, an attempt by Bowcock et. al. \cite{Ref2} which starts from
brane-based approach has also given the same result.

Our aim in this paper is, to construct a general formalism for coordinate
transformation between the bulk-based and the brane-based approaches and
then investigating their equivalence in the case of anisotropic models.
Observations at angular distribution of extra-galactic radio sources,
spatial distribution of the redshifts of extra-galactic objects and
temperature distribution of cosmic microwave radiation confirm that our
universe has anisotropy, so it is worth to investigate to anisotropic models
to get a more realistic description of the universe. In Sec.II, applying
transformation method in \cite{Ref1}, we obtain the transformed brane-based
metric from the most general bulk-based anisotropic metric ansatz. After
obtaining the solution of bulk field equations, we apply the method for a
particular anisotropic model in Sec.III and finally Sec.IV is devoted to the
summary of the paper and discussions.

\section{Transformation from the bulk-based approach to brane-based one}

The most general 5-dimensional anisotropic bulk-based metric\footnote[1]{%
Although the most general anisotropic metric has non-diagonal form for every
components, it becomes (\ref{bulkmetric}) in the case of providing non-zero
Killing field requirement.} admitting non-zero Killing vector space, which
preserves symmetries along congurence lines is 
\begin{equation}
ds^{2}=-A_{0}(\hat{r})d\hat{t}^{2}+A_{ij}(\hat{r})d\hat{x}^{i}d\hat{x}%
^{j}+A_{4}(\hat{r})d\hat{r}^{2}  \label{bulkmetric}
\end{equation}%
where $\hat{r}$ denotes the extra spatial coordinate and $i,j=1,2,3$ are
indices of 3-dimensional spacetime. The 4-dimensional brane, corresponding
to our universe moves along the extra dimension and is described by $(\tau
,x^{i})$ coordinates. Base vectors and one forms, in the 5-dimensional and
4-dimensional spacetime, respectively are given in below. 
\begin{equation}
e_{A}\equiv \partial _{A}=\frac{\partial }{\partial \hat{x}_{A}}=(\partial _{%
\hat{t}},\partial _{\hat{x}^{i}},\partial _{\hat{r}})\omega ^{A}\equiv d\hat{%
x}^{A}=(d\hat{t},d\hat{x}^{i},d\hat{r})
\end{equation}%
\begin{equation}
e_{\mu }\equiv \partial _{\mu }=\frac{\partial }{\partial \hat{x}_{\mu }}%
=(\partial _{\tau },\partial _{x^{i}})\omega ^{\mu }\equiv dx^{\mu }=(d\tau
,dx^{i})
\end{equation}%
Here we define $A=0..4$, $\mu =0..3.$ The brane represented by $\hat{r}=R(%
\hat{t})$ hypersurfaces can be induced on 4-dimensional spacetime via the
following transformations.

\begin{equation}
\begin{array}{c}
\hat{t}=T(\tau )\rightarrow d\hat{t}=\dot{T}d\tau \\ 
\hat{x}^{i}=x^{i}\rightarrow d\hat{x}^{i}=dx^{i} \\ 
\hat{r}=R(\tau )\rightarrow d\hat{r}=\dot{R}d\tau%
\end{array}%
\end{equation}

The induced metric on brane is then 
\begin{equation}
ds_{brane}^{2}=-(A_{0}\dot{T}^{2}-A_{4}\dot{R}^{2})d\tau ^{2}+A_{ij}(R(\tau
))dx^{i}dx^{j}
\end{equation}
where we introduced cosmological time $\tau $ and cosmological scale factor $%
R(\tau )$. The dot denotes the derivative respect to $\tau $. Now we can
construct a vector space generated by tangent vectors of geodesics
intersecting with hypersurface $\hat{r}=R(\hat{t})$ perpendicularly 
\begin{equation}
u^{A}=e_{\tau }^{A}\partial _{A}=\dot{T}\partial _{\hat{t}}+\dot{R}\partial
_{\hat{r}}  \label{tangent1}
\end{equation}

We choose geodesics as spacelike and having zero $\hat{x}^{i}$-components to
provide a timelike hypersurface. The Killing field of bulk spacetime helps
us to find constants of motion along geodesics: 
\begin{equation}
g_{AB}u^{A}\xi ^{B}=-E
\end{equation}%
\begin{equation}
g_{AB}u^{A}u^{B}=1
\end{equation}%
where $E$ is an constant of integration. Using tangent vector's components
in (\ref{tangent1}), we obtain 
\begin{equation}
u^{A}=\left( \frac{E}{A_{0}},0,0,0,\mp \frac{A_{0}+E^{2}}{A_{0}A_{4}}\right)
\label{comp}
\end{equation}

The trajectory of the geodesic is given by 
\begin{equation}
\frac{dx^{A}}{dw}=u^{A}  \label{traj}
\end{equation}%
where $w$ is the affine parameter. All points ($P$) on the hypersurface
described by $(\tau ,x^{i})$ coordinates intersecting perpendicularly with
an affinely parameterized geodesic. Hence we can describe the point $P$,
with a new coordinate set $(\tau ,x^{i},w)$, where the new coordinate $w$ is
now an extra spatial coordinate of $P$ and this system is called brane-based
coordinates. One can easily construct the brane-based metric, from the
bulk-based one by applying transformations: $\hat{r}=\hat{r}(\tau ,w)$, $%
\hat{t}=\hat{t}(\tau ,w)$ 
\begin{equation}
d\hat{t}=\left( \frac{\partial \hat{t}}{\partial \tau }\right) d\tau +\left( 
\frac{\partial \hat{t}}{\partial w}\right) dw=e_{\tau }^{\hat{t}}d\tau
+e_{w}^{\hat{t}}dw
\end{equation}%
\begin{equation}
d\hat{r}=\left( \frac{\partial \hat{r}}{\partial \tau }\right) d\tau +\left( 
\frac{\partial \hat{r}}{\partial w}\right) dw=e_{\tau }^{\hat{r}}d\tau
+e_{w}^{\hat{r}}dw
\end{equation}%
Substituting them in (\ref{bulkmetric}) gives 
\begin{eqnarray}
ds^{2} &=&-\left[ A_{0}\left( \frac{\partial \hat{t}}{\partial \tau }\right)
^{2}-A_{4}\left( \frac{\partial \hat{r}}{\partial \tau }\right) ^{2}\right]
d\tau ^{2}+A_{ij}dx^{i}dx^{j}  \notag \\
&&+\left[ -A_{0}\left( \frac{\partial \hat{t}}{\partial w}\right)
^{2}+A_{4}\left( \frac{\partial \hat{r}}{\partial w}\right) ^{2}\right]
dw^{2}  \notag \\
&&+2\left[ -A_{0}\left( \frac{\partial \hat{t}}{\partial \tau }\right)
\left( \frac{\partial \hat{t}}{\partial w}\right) +A_{4}\left( \frac{%
\partial \hat{r}}{\partial \tau }\right) \left( \frac{\partial \hat{r}}{%
\partial w}\right) \right] d\tau dw  \label{transformed}
\end{eqnarray}

The final metric in (\ref{transformed}) is a general form of brane-based
metric which is transformed from the bulk-based metric. We need to find the
exact forms of transformation coefficients denoted by partial derivatives in
(\ref{transformed}). Replacing (\ref{comp}) into (\ref{traj}) we get two
equations 
\begin{equation}
u^{\hat{t}}=e_{w}^{\hat{t}}=\frac{\partial \hat{t}}{\partial w}=\frac{E}{%
A_{0}}  \label{relation1}
\end{equation}%
\begin{equation}
u^{\hat{r}}=e_{w}^{\hat{r}}=\frac{\partial \hat{r}}{\partial w}=\mp \sqrt{%
\frac{A_{0}+E^{2}}{A_{0}A_{4}}}  \label{relation}
\end{equation}%
of which the second one gives an integral relation between extra coordinates
of the two approaches. In the case the components of the bulk-based metric
are known, (\ref{relation}) can be solved exactly. 
\begin{equation}
\mp w+w_{0}(\tau )=\int \frac{d\hat{r}}{\sqrt{\frac{A_{0}+E^{2}}{A_{0}A_{4}}}%
}  \label{intg}
\end{equation}

On the other hand, we get transverse coefficients in (\ref{relation1}) and (%
\ref{relation}) from $dw/dx^{B}=g_{AB}u^{A}$ 
\begin{equation}
e_{\hat{t}w}=\frac{\partial w}{\partial \hat{t}}=-E
\end{equation}
\begin{equation}
e_{\hat{x}^{i}w}=\frac{\partial w}{\partial \hat{x}^{i}}=0
\end{equation}
\begin{equation}
e_{\hat{r}w}=\frac{\partial w}{\partial \hat{r}}=\pm \sqrt{\frac{A_{4}\left(
A_{0}+E^{2}\right) }{A_{0}}}
\end{equation}

The integrability condition $ddw=0$ is equivalent to 
\begin{equation}
\left( \frac{\partial w}{\partial \hat{t}}\right) d\hat{t}=-\left( \frac{%
\partial w}{\partial \hat{r}}\right) d\hat{r}
\end{equation}%
and gives a ratio of coefficients 
\begin{equation}
\left( \frac{\partial \hat{t}}{\partial \tau }\right) /\left( \frac{\partial 
\hat{r}}{\partial \tau }\right) =-\left( \frac{\partial w}{\partial \hat{r}}%
\right) /\left( \frac{\partial w}{\partial \hat{t}}\right) =\pm \sqrt{\frac{%
A_{4}\left( A_{0}+E^{2}\right) }{A_{0}E^{2}}}
\end{equation}%
The transverse ratio is then 
\begin{equation}
\left( \frac{\partial \tau }{\partial \hat{t}}\right) /\left( \frac{\partial
\tau }{\partial \hat{r}}\right) =-\left( \frac{\partial \hat{r}}{\partial w}%
\right) /\left( \frac{\partial \hat{t}}{\partial w}\right) =\pm \sqrt{\frac{%
A_{0}\left( A_{0}+E^{2}\right) }{A_{4}E^{2}}}
\end{equation}

In order to see all transformation coefficients which have been calculated
up to now, let us put them into a matrix form as below 
\begin{equation}
(\hat{t},\hat{x}^{i},\hat{r})\rightarrow (\tau ,x^{i},w) : \left( 
\begin{array}{ccc}
e_{\tau }^{\hat{t}} & e_{x^{i}}^{\hat{t}} & e_{w}^{\hat{t}} \\ 
e_{\tau }^{\hat{x}^{i}} & e_{x^{i}}^{\hat{x}^{i}} & e_{w}^{\hat{x}^{i}} \\ 
e_{\tau }^{\hat{r}} & e_{x^{i}}^{\hat{r}} & e_{w}^{\hat{r}}%
\end{array}
\right) =\left( 
\begin{array}{ccc}
a_{11} & 0 & \frac{E}{A_{0}} \\ 
0 & 1 & 0 \\ 
a_{31} & 0 & \pm \sqrt{\frac{\left( A_{0}+E^{2}\right) }{A_{4}A_{0}}}%
\end{array}
\right)
\end{equation}
\begin{equation}
(\tau ,x^{i},w)\rightarrow (\hat{t},\hat{x}^{i},\hat{r}) : \left( 
\begin{array}{ccc}
e_{\hat{t}\tau }^{{}} & e_{\hat{t}x^{i}}^{{}} & e_{\hat{t}w}^{{}} \\ 
e_{\hat{x}^{i}\tau }^{{}} & e_{\hat{x}^{i}x^{i}}^{{}} & e_{\hat{x}^{i}w}^{{}}
\\ 
e_{\hat{r}\tau }^{{}} & e_{\hat{r}x^{i}}^{{}} & e_{\hat{r}w}^{{}}%
\end{array}
\right) =\left( 
\begin{array}{ccc}
a^{11} & 0 & -E \\ 
0 & 1 & 0 \\ 
a^{31} & 0 & \pm \sqrt{\frac{A_{4}\left( A_{0}+E^{2}\right) }{A_{0}}}%
\end{array}
\right)
\end{equation}
Note that, two of the matrix components remain unknown, but they depend on
each other according to these equations 
\begin{equation}
e_{\tau }^{\hat{t}}/e_{\tau }^{\hat{r}}=a_{11}/a_{31}=\pm \sqrt{\frac{
A_{4}\left( A_{0}+E^{2}\right) }{A_{0}E^{2}}}
\end{equation}
\begin{equation}
e_{\hat{t}\tau }^{{}}/e_{\hat{r}\tau }^{{}}=a^{11}/a^{31}=\pm \sqrt{\frac{
A_{0}\left( A_{0}+E^{2}\right) }{A_{4}E^{2}}}
\end{equation}
Substituting partial derivatives from the matrix in (\ref{transformed}), we
find the final form of the brane-based metric 
\begin{equation}
ds^{2}=-\frac{A_{0}(\hat{r}(\tau ,w))A_{4}(\hat{r}(\tau ,w))}{E^{2}}\left( 
\frac{\partial \hat{r}}{\partial \tau }\right) ^{2}d\tau ^{2}+A_{ij}(\hat{r}
(\tau ,w))dx^{i}dx^{j}+dw^{2}\   \label{lastmetric}
\end{equation}
which is a general form of the brane-based metric in Gaussian Normal
Coordinates and it contains spatial anisotropy as expected. If we choose
metric components as $A_{0}=f(\hat{r})$, $A_{4}=1/f(\hat{r})$ and the
3-dimensional metric as isotropic, Eq. (\ref{lastmetric}) readily yields the
result in Ref. \cite{Ref1}.

All terms in (34) depend on brane coordinates $(\tau ,w)$. The motion
constant $E$ of bulk geodesics is not a constant anymore on the brane and
turns out to be $E=E(\tau )$. Besides, the term showing partial derivative
can be calculated from integral relation (\ref{intg}) after obtaining
explicit forms of metric components. For this purpose one must solve vacuum
Einstein field equations in the bulk and replace them with the transformed
metric. (\ref{lastmetric}) gives the induced metric on brane by setting $w=0$%
.

\section{\protect\bigskip Applications of the Method for an Anisotropic
Bulk-Based Model}

To demonstrate an example of the above transformation, we choose the
bulk-based metric below 
\begin{equation}
ds^{2}=-f(\hat{r})d\hat{t}^{2}+\hat{r}^{2}(a(\hat{r})d\hat{x}^{2}+b(\hat{r})d%
\hat{y}^{2}+c(\hat{r})d\hat{z}^{2})+\frac{1}{f(\hat{r})}d\hat{r}^{2}
\label{assumption}
\end{equation}%
which still conserves spatial anisotropy with three different diagonal terms 
$a(\hat{r})$, $b(\hat{r})$ and $c(\hat{r})$. Our assumption is diversed from
the one in \cite{Ref28} with the factor of dependence of functions'
variables. 5-dimensional vacuum Einstein field equations\footnote[2]{%
Field equations are given in the appendix.} give two equations for $f(r)$,
which is a remarkably important term to find the brane-based metric by
transformations given in the previous section. 
\begin{equation}
f^{\prime \prime }+\frac{f^{\prime }}{2}\left( \frac{a^{\prime }}{a}+\frac{%
b^{\prime }}{b}+\frac{c^{\prime }}{c}\right) =\frac{4\Lambda _{5}}{3}
\label{f1}
\end{equation}%
\begin{equation}
\left( \frac{f^{\prime }}{4}+\frac{f}{\hat{r}}\right) \left( \frac{a^{\prime
}}{a}+\frac{b^{\prime }}{b}+\frac{c^{\prime }}{c}\right) +\frac{f}{4}\left( 
\frac{a^{\prime }}{a}\frac{b^{\prime }}{b}+\frac{a^{\prime }}{a}\frac{%
c^{\prime }}{c}+\frac{b^{\prime }}{b}\frac{c^{\prime }}{c}\right) +\frac{%
3f^{\prime }}{2\hat{r}}+\frac{3f}{\hat{r}^{2}}=\Lambda _{5}  \label{f2}
\end{equation}%
Defining $p,q,k$ as constants, we can get an exact solution of $f(r)$ for
particular functions 
\begin{equation}
a(\hat{r})=\hat{r}^{p},b(\hat{r})=\hat{r}^{q},c(\hat{r})=\hat{r}^{k}
\end{equation}%
and equations (\ref{f1}, \ref{f2}) give rise to 
\begin{equation}
f(\hat{r})=\frac{4\Lambda _{5}}{3(p+q+k+2)}\hat{r}^{2}-C
\end{equation}%
with the condition $p+q+k=6.$ Here $C$ is the constant of integration. Then (%
\ref{assumption}) arises 
\begin{equation}
ds^{2}=-\left( \frac{\Lambda _{5}}{6}\hat{r}^{2}+C\right) d\hat{t}^{2}+\hat{r%
}^{2}(\hat{r}^{p}d\hat{x}^{2}+\hat{r}^{q}d\hat{y}^{2}+\hat{r}^{k}d\hat{z}%
^{2})+\left( \frac{\Lambda _{5}}{6}\hat{r}^{2}+C\right) ^{-1}d\hat{r}^{2}\ 
\label{adstype}
\end{equation}%
\newline
as in the form of AdS spacetime with $\Lambda _{5}=-\frac{6}{l^{2}}$. Note
that this metric differs from AdS spacetime by its anisotropic spatial
section. Substituting $A_{0}=f(\hat{r})=\frac{\Lambda _{5}}{6}\hat{r}^{2}+C$
and $A_{4}=1/f(\hat{r})$ in (\ref{intg}) 
\begin{equation}
\mp w+w_{0}(\tau )=\int \frac{d\hat{r}}{\sqrt{\frac{A_{0}+E^{2}}{A_{0}A_{4}}}%
}=\int \frac{d\hat{r}}{\sqrt{\frac{\Lambda _{5}}{6}\hat{r}^{2}+C+E^{2}}}
\end{equation}%
we get the relation for two extra coordinates as follows 
\begin{equation}
\mp w+w_{0}(\tau )=\left\{ 
\begin{array}{c}
\frac{6}{\Lambda _{5}}\sinh ^{-_{1}}\left( \sqrt{\frac{6}{\Lambda _{5}}}%
\frac{\hat{r}}{C+E^{2}}\right) ,\qquad C+E^{2}>1 \\ 
\\ 
~\frac{6}{\Lambda _{5}}\sin {}^{-_{1}}\left( \sqrt{\frac{6}{\Lambda _{5}}}%
\frac{\hat{r}}{C+E^{2}}\right) ,\qquad C+E^{2}<1%
\end{array}%
\right.
\end{equation}%
Using the fact that geodesics intersect with the hypersurface $%
r(t_{0})=R(t_{0}(\tau ))$ where $w=0$ at $t=t_{0}$, we define $w_{0}(\tau )$
and then get 
\begin{equation}
\hat{r}(\tau ,w)=\left\{ 
\begin{array}{c}
Q_{+}(\tau )\sinh \left( \pm \sqrt{\frac{\Lambda _{5}}{6}}w\right) +R(\tau
)\cosh \left( \pm \sqrt{\frac{\Lambda _{5}}{6}}w\right) ,\qquad C+E^{2}>1 \\ 
\\ 
Q_{-}(\tau )\sin \left( \pm \sqrt{\frac{\Lambda _{5}}{6}}w\right) +R(\tau
)\cos \left( \pm \sqrt{\frac{\Lambda _{5}}{6}}w\right) ,\qquad C+E^{2}<1%
\end{array}%
\right.
\end{equation}%
\newline
where we represent 
\begin{equation}
E=\dot{R}(\tau )
\end{equation}%
\begin{equation}
Q_{\pm }(\tau )=\sqrt{\frac{\Lambda _{5}}{6}\pm \frac{R(\tau )}{C+\dot{R}%
^{2}(\tau )}}\left[ C+\dot{R}^{2}(\tau )\right]
\end{equation}%
Calculating partial derivative in (\ref{lastmetric}), we find the
corresponding brane-based metric of our anisotropic assumption in (\ref%
{assumption}) 
\begin{equation}
ds^{2}=-\Phi (\tau ,w)d\tau ^{2}+\hat{r}^{2}(\tau ,w)\left[ \hat{r}^{p}(\tau
,w)d{x}^{2}+\hat{r}^{q}(\tau ,w)d{y}^{2}+\hat{r}^{k}(\tau ,w)d{z}^{2}\right]
+dw^{2}  \label{assumpresult}
\end{equation}%
\newline
Here we represent $\Phi (\tau ,w)$ as 
\begin{equation}
\Phi (\tau ,w)=\left\{ 
\begin{array}{c}
\frac{1}{HR}\cosh \left( \pm \sqrt{\frac{\Lambda _{5}}{6}}w\right) +\varphi
(\tau )\sinh \left( \pm \sqrt{\frac{\Lambda _{5}}{6}}w\right) ,\qquad
C+E^{2}>1 \\ 
\\ 
\frac{1}{HR}\cos \left( \pm \sqrt{\frac{\Lambda _{5}}{6}}w\right) +\varphi
(\tau )\sin \left( \pm \sqrt{\frac{\Lambda _{5}}{6}}w\right) ,\qquad
C+E^{2}<1%
\end{array}%
\right.
\end{equation}%
and 
\begin{equation}
\varphi (\tau )=\frac{\frac{4\Lambda _{5}}{6}\left( H^{2}R^{2}+C\right)
\left( H^{2}+\dot{H}\right) \pm CR^{-1}\pm R\left( 3H^{2}+\dot{H}\right) }{%
\sqrt{\frac{\Lambda _{5}}{6}\left( C\left( RH\right) ^{-1}+RH\right) ^{2}\pm
R\left( 1+\left( RH\right) ^{-2}\right) }}^{2}
\end{equation}%
$H$ denotes $\dot{R}(\tau )/R$ . Here we see that depending on the positive
(or negative) values of $C$, geometry of the Universe will be open (or
closed) with hyperbolic (or elliptic) evolution. Metric (\ref{assumpresult})
is completely written in the brane-based coordinates and has the form of
Gaussian Normal Coordinates. Additionally it is possible to derive the
metric on the brane by setting $w=0$ as below 
\begin{equation}
ds^{2}=-\frac{1}{HR}d\tau ^{2}+R^{2}(\tau )\left[ R^{p}(\tau )d{x}%
^{2}+R^{q}(\tau )d{y}^{2}+R^{k}(\tau )d{z}^{2}\right]  \label{ended}
\end{equation}%
which is in the form of Kasner-like spacetime.

\section{\protect\bigskip Discussion}

We have studied 5-dimensional brane-world models in the context of two
different approaches which are widely used for brane-world cosmologies. One
can always build a coordinate transformation linking these approaches under
the condition that spacetime symmetries exist, i.e., non-zero Killing vector
fields. Although it is possible to constitute a special transformation for
each metric respectively, it will certainly be useful to construct a
operable transformation procedure. We have first tried to rewrite such a
generalized procedure through the formalism presented in \cite{Ref1}.

Starting from the most general 5-dimensional bulk-based metric with
homogeneous and anisotropic spatial 3-sections (\ref{bulkmetric}), we have
shown all steps of transformation, keeping metric components nonspecific in
order to make any particular metric applicable. One of the most important
features of the transformation is the integral equation (\ref{intg}) which
relates coordinates of extra dimensions of the two approaches. When metric
components are determined from field equations, this equation can be solved
and the transformed metric (\ref{lastmetric}) becomes explicit.

It is also interesting that the transformed metric (\ref{lastmetric}) is in
the form of Gaussian Normal coordinates. However anisotropy in the spatial
sections remains formally unaffected under the transformation and can be
obtained exactly from the solution of (\ref{intg}).

In order to employ specific metric components in the transformation, we then
derived the solution of a more restricted but anisotropic spacetime (\ref%
{assumption}) allowing analytical calculations of the bulk field equations.
Our solution (\ref{adstype}) differs from AdS spacetime only with the
anisotropic 3-spatial sections, however it is clear that it represents
conformally AdS spacetime. Defining $k={\Lambda _{5}}/{6}$ and setting $C=1$%
, $\hat{x}$, $\hat{y}$, $\hat{z}=constant$, 2-dimensional conformal
spacetime will be

\begin{equation}
ds^{2}=-\left( k^{2}\hat{r}^{2}+1\right) d\hat{t}^{2}+\frac{1}{\left( k^{2}%
\hat{r}^{2}+1\right) }d\hat{r}^{2}
\end{equation}%
which can also be regarded as a BTZ black hole in the bulk \cite{Ref29}.

After performing transformation for the chosen metric, we have obtained the
brane-based version of (\ref{adstype}) as given in (\ref{assumpresult}). The
cosmological behavior changes depending on values of $C+E^{2}$ being either
elliptic or hyperbolic. It is worth emphasizing that the induced metric on
the brane (\ref{ended}) represents the Kasner type spacetime when the fifth
coordinate is fixed to zero. However one needs further calculations for the
scale factor $R(\tau )$, which requires initial conditions and the
matter-energy distribution on the brane have been determined.

\section{Acknowledgements}

\appendix

\section{Field Equations for the Anisotropic Bulk-based Model}

\begin{equation}
\frac{f}{2}\left( \frac{a^{\prime \prime }}{a}+\frac{b^{\prime \prime }}{b}+%
\frac{c^{\prime \prime }}{c}\right) +\left( \frac{f^{\prime }}{4}+\frac{2f}{%
\hat{r}}\right) \left( \frac{a^{\prime }}{a}+\frac{b^{\prime }}{b}+\frac{%
c^{\prime }}{c}\right) -\frac{f}{4}\left( \left( \frac{a^{\prime }}{a}%
\right) ^{2}+\left( \frac{b^{\prime }}{b}\right) ^{2}+\left( \frac{c^{\prime
}}{c}\right) ^{2}\right) +\frac{f}{4}\left( \frac{a^{\prime }}{a}\frac{%
b^{\prime }}{b}+\frac{a^{\prime }}{a}\frac{c^{\prime }}{c}+\frac{b^{\prime }%
}{b}\frac{c^{\prime }}{c}\right) +\frac{3f^{\prime }}{2\hat{r}}+\frac{3f}{%
\hat{r}^{2}}=\Lambda _{5}
\end{equation}

\begin{equation}
\frac{f}{2}\left( \frac{b^{\prime \prime }}{b}+\frac{c^{\prime \prime }}{c}%
\right) +\left( \frac{f^{\prime }}{2}+\frac{3f}{2\hat{r}}\right) \left( 
\frac{b^{\prime }}{b}+\frac{c^{\prime }}{c}\right) -\frac{f}{4}\left( \left( 
\frac{b^{\prime }}{b}\right) ^{2}+\left( \frac{c^{\prime }}{c}\right)
^{2}\right) +\frac{f}{4}\left( \frac{b^{\prime }}{b}\frac{c^{\prime }}{c}%
\right) +\frac{f^{\prime \prime }}{2}+\frac{f^{\prime }}{2\hat{r}}+\frac{f}{%
\hat{r}^{2}}=\Lambda _{5}
\end{equation}

\begin{equation}
\frac{f}{2}\left( \frac{a^{\prime \prime }}{a}+\frac{c^{\prime \prime }}{c}%
\right) +\left( \frac{f^{\prime }}{2}+\frac{3f}{2\hat{r}}\right) \left( 
\frac{a^{\prime }}{a}+\frac{c^{\prime }}{c}\right) -\frac{f}{4}\left( \left( 
\frac{a^{\prime }}{a}\right) ^{2}+\left( \frac{c^{\prime }}{c}\right)
^{2}\right) +\frac{f}{4}\left( \frac{a^{\prime }}{a}\frac{c^{\prime }}{c}%
\right) +\frac{f^{\prime \prime }}{2}+\frac{f^{\prime }}{2\hat{r}}+\frac{f}{%
\hat{r}^{2}}=\Lambda _{5}
\end{equation}

\begin{equation}
\frac{f}{2}\left( \frac{a^{\prime \prime }}{a}+\frac{b^{\prime \prime }}{b}%
\right) +\left( \frac{f^{\prime }}{2}+\frac{3f}{2\hat{r}}\right) \left( 
\frac{a^{\prime }}{a}+\frac{b^{\prime }}{b}\right) -\frac{f}{4}\left( \left( 
\frac{a^{\prime }}{a}\right) ^{2}+\left( \frac{b^{\prime }}{b}\right)
^{2}\right) +\frac{f}{4}\left( \frac{a^{\prime }}{a}\frac{b^{\prime }}{b}%
\right) +\frac{f^{\prime \prime }}{2}+\frac{f^{\prime }}{2\hat{r}}+\frac{f}{%
\hat{r}^{2}}=\Lambda _{5}
\end{equation}

\begin{equation}
\left( \frac{f^{\prime }}{4}+\frac{f}{\hat{r}}\right) \left( \frac{a^{\prime
}}{a}+\frac{b^{\prime }}{b}+\frac{c^{\prime }}{c}\right) +\frac{f}{4}\left( 
\frac{a^{\prime }}{a}\frac{b^{\prime }}{b}+\frac{a^{\prime }}{a}\frac{%
c^{\prime }}{c}+\frac{b^{\prime }}{b}\frac{c^{\prime }}{c}\right) +\frac{%
3f^{\prime }}{2\hat{r}}+\frac{3f}{\hat{r}^{2}}=\Lambda _{5}
\end{equation}%
Equations correspond $(00)$, $(ii)$, $(44)$ respectively. The prime denotes
the differentiate with respect to $\hat{r}$.

\end{document}